# Summarizing Semantic Model Differences

Shahar Maoz*, Jan Oliver Ringert**, and Bernhard Rumpe

Software Engineering
RWTH Aachen University, Germany
http://www.se-rwth.de/

**Abstract.** Fundamental building blocks for managing and understanding software evolution in the context of model-driven engineering are differencing operators one can use for model comparisons. Semantic model differencing deals with the definition and computation of semantic diff operators for model comparison, operators whose input consists of two models and whose output is a set of diff witnesses, instances of one model that are not instances of the other. However, in many cases the complete set of diff witnesses is too large to be efficiently computed and effectively presented. Moreover, many of the witnesses are very similar and hence not interesting. Thus, an important challenge of semantic differencing relates to witness selection and presentation.
In this paper we propose to address this challenge using a summarization technique, based on a notion of equivalence that partitions the set of diff witnesses. The result of the computation is a summary set, consisting of a single representative witness from each equivalence class. We demonstrate our ideas using two concrete diff operators, for class diagrams and for activity diagrams, where the computation of the summary set is efficient and does not require the enumeration of all witnesses.

## 1 Introduction

Differencing operators used for model comparisons are fundamental building blocks for managing and understanding software evolution in model-driven engineering. Semantic model differencing [12] deals with the definition and computation of semantic diff operators, whose input consists of two models, e.g., two versions along the history of a model, and whose output is a set of diff witnesses, instances of one model that are not instances of the other. Each witness serves as a concrete proof for the difference between the two models and its meaning.

However, the complete set of diff witnesses is in many cases too large to be efficiently computed and effectively presented. Moreover, many of the witnesses are very similar and hence not interesting. Thus, an important challenge of semantic differencing relates to witness computation, selection, and presentation.

In this paper we propose to address this challenge using the definition of a summarization technique, based on a notion of equivalence that partitions the set of diff witnesses. The result of the summarization is a summary set, consisting of a single representative witness from each equivalence class.

---

* S. Maoz acknowledges support from a postdoctoral Minerva Fellowship, funded by the German Federal Ministry for Education and Research.
** J.O. Ringert is supported by the DFG GK/1298 AlgoSyn.

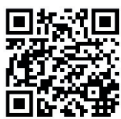

In recent work we have presented two concrete semantic diff operators, *cddiff* [11] for class diagrams (CDs) and *addiff* [9] for activity diagrams (ADs), along with the algorithms to compute them and with an initial evaluation of their performance and the usefulness of their results. Here we demonstrate the application of the summarization technique to these two concrete diff operators. Moreover, the computation of the summary set is efficient and does not require the enumeration of all witnesses.

It is important to note that we do not look for a difference summary in the form of a succinct mathematical representation of all differences between the two models, e.g., in the case of activity diagrams, a state machine accepting exactly all those traces accepted by one model and not by the other. Rather, we believe that in order to make semantic differencing useful and attractive to engineers, one needs to take a constructive and concrete approach: to compute and present concrete, specific, and thus easy to understand witnesses for the difference (e.g., in the case of activity diagrams, concrete execution traces).

Sect. 2 presents examples to motivate the need for summarization. Sect. 3 presents a formal, language independent overview of our approach and continues with its specializations for CDs and ADs. Sect. 4 briefly describes the algorithms used to compute the summarized sets of witnesses. Initial evaluation and discussion appear in Sect. 5. Related work is discussed in Sect. 6 and Sect. 7 concludes.

## 2 Examples

**Example I.** Consider $cd.v1$ of Fig. 1, describing a first version of a model for (part of) a company structure with employees, managers, and tasks. A design review with a domain expert has revealed three bugs in this model: (1) the number of tasks per employee should not be limited to two; (2) managers are also employees, and they can handle tasks too; (3) an employee must have exactly one manager. These bugs have been addressed in the second version $cd.v2$.

Diff witnesses for the semantic difference between $cd.v2$ and $cd.v1$ are object models that are in the semantics of $cd.v2$ and not in the semantics of $cd.v1$. Fig. 2 shows two such diff witnesses: $om_1$, consisting of an employee with three tasks, who is managed by a manager; and $om_2$, consisting of a manager that manages herself, without any tasks. However, these are only examples. Many more diff witnesses exist, e.g., those that are similar to $om_1$ but include additional tasks, or those that consist of duplicates of $om_1$ and/or $om_2$ etc.

**Example II.** Consider the ADs of Fig. 3, describing three versions of a ticket reservation process. Witnesses for the semantic difference between two ADs are execution traces that are allowed by one AD and are not allowed by the other.

For example, traces of $ad.v2$ that are not in $ad.v1$ include (1) a trace with $tickets = 3$ where the action `accounts` comes before the action `reserve`, and (2) a trace with $tickets = 10$ (where $ad.v2$ executes actions `register` and `welcome msg`). Traces of $ad.v3$ that are not in $ad.v2$ are all traces with $tickets < 12$.

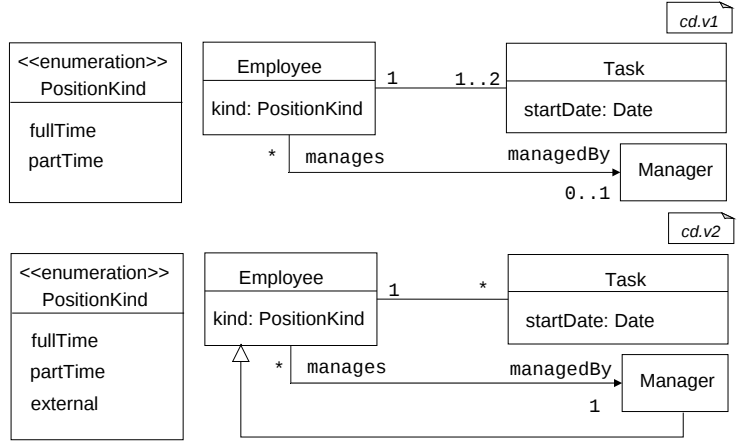

**Fig. 1.** Two versions of a CD, $cd.v1$ and $cd.v2$.

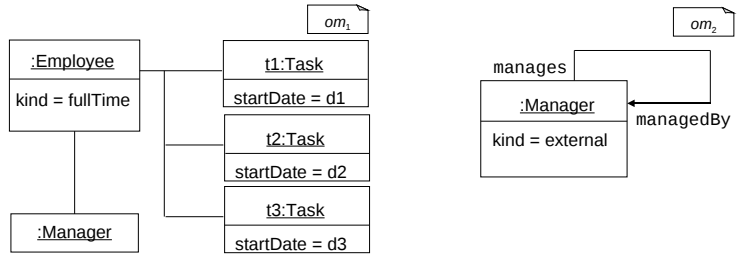

**Fig. 2.** Two diff witnesses from $cddiff(cd.v2, cd.v1)$.

Overall, there are many diff traces, due to the possible values of the input *tickets* and the partial order between `reserve`, `accounts`, and `updates`.

Given the large number of diff witnesses, for both *cddiff* and *addiff*, the challenge we address in this paper relates to the computation, selection, and presentation of a summarized set of witnesses.

## 3 Definitions

Consider a modeling language $ML = \langle Syn, Sem, sem \rangle$ where $Syn$ is the set of all syntactically correct expressions (models) according to some syntax definition, $Sem$ is a semantic domain, and $sem : Syn \to \mathcal{P}(Sem)$ is a function mapping each expression $e \in Syn$ to a set of elements from $Sem$ (see [5]).

The semantic diff operator $\mathit{diff} : Syn \times Syn \to \mathcal{P}(Sem)$ maps two syntactically correct expressions $e_1$ and $e_2$ to the (possibly infinite) set of all $s \in Sem$ that are in the semantics of $e_1$ and not in the semantics of $e_2$. Formally:

**Definition 1.** $\mathit{diff}(e_1, e_2) = \{s \in Sem |\ s \in sem(e_1) \land s \notin sem(e_2)\}$.

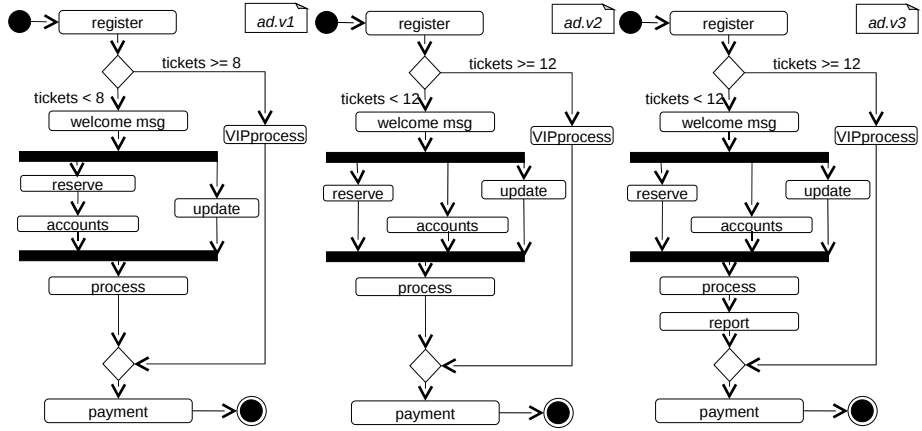

**Fig. 3.** Three versions of an AD for a ticket reservation process. The input variable *tickets* ranges from 0 to 15.

The elements in $\mathit{diff}(e_1, e_2)$ are called *diff witnesses*. When $e_1$ and $e_2$ are fixed, we use *diff* for the set $\mathit{diff}(e_1, e_2)$.

Let $Q = \{Q_1, Q_2, \ldots\}$ be a partition of *diff*, that is, $\mathit{diff} = \bigcup Q_i$, $\forall i : Q_i \neq \emptyset$ and $\forall i \neq j : Q_i \cap Q_j = \emptyset$. We define a partition function $\mathit{part} : \mathit{diff} \to Q$, which maps every diff witness $dw \in \mathit{diff}$ to an element $Q_i$ of the partition $Q$ such that $dw \in Q_i$. Note that $Q$, *diff*, and *part* all depend on fixed $e_1$ and $e_2$.

A summary of the set *diff* according to a partition $Q$, $\mathit{diff}_Q$, is a subset of *diff* consisting of a representative diff witness from each element in $Q$. Formally:

**Definition 2.** *Given a set of diff witnesses $\mathit{diff} = \mathit{diff}(e_1, e_2)$ and a partition $Q$, a summary of diff is a set $\mathit{diff}_Q \subseteq \mathit{diff}$ s.t.*

1. $\forall dw_1, dw_2 \in \mathit{diff}_Q$, $dw_1 \neq dw_2 \Rightarrow \mathit{part}(dw_1) \neq \mathit{part}(dw_2)$
2. $\forall Q_i \in Q$, $\exists dw \in \mathit{diff}_Q$ *s.t.* $\mathit{part}(dw) = Q_i$.

### 3.1 Specialization for class diagrams

In previous work we have defined *cddiff*, a specializations of *diff* for CDs [11]. We now present a related specialization of $\mathit{diff}_Q$.

Our semantics of CDs is based on [4] and is given in terms of sets of objects and their relationships. Thus, the elements of *cddiff* are object models (and they are presented to the engineer using object diagrams). To define $\mathit{cddiff}_Q$, we define a partition of the set of all object models $\mathit{cddiff} \subseteq OM$ into equivalence classes based on the set of classes instantiated in each object model. More formally:

**Definition 3 (class-equivalent partition for object models).** *The class-equivalent partition maps every object model $om_1 \in \mathit{cddiff}$ to the set of all object models in cddiff whose set of instantiated classes is equal to the set of classes instantiated in $om_1$: $\mathit{part}(om_1) = \{om \mid \mathit{classes}(om) = \mathit{classes}(om_1)\}$.*

For example, consider the CDs shown in Sect. 2. A summary of the semantic difference $cddiff(cd.v2, cd.v1)$, according to the class equivalence partition, will include exactly four object models: one consisting only of managers (an example representative is a single manager managing herself, with no tasks and no employees that are not managers), another one consisting of only managers and employees (an example representative is a manager who manages an employee with no tasks), another one with only managers and tasks (an example representative is a manager managing herself and having several tasks), and, finally, one consisting of managers, employees, and tasks (an example representative is a model consisting of a manager managing an employee with several tasks).

### 3.2 Specialization for activity diagrams

In previous work we have defined *addiff*, a specializations of *diff* for ADs [9]. We now present related specializations of $diff_Q$.

We use UML2 Activity Diagrams for the syntax of our ADs. In addition to action nodes, pseudo nodes (fork, decision, etc.), the language includes input and local variables (over finite domains), transition guards, and assignments. Roughly, the semantics of an AD is made of a set of finite action traces starting from an initial node, considering interleaving execution of fork branches, the guards on decision nodes etc. (a formal and complete semantics of our ADs is given in [10]). The elements of *addiff* are execution traces of one AD that are not possible in the other AD; we call them diff traces. Note that we do not require that the traces end in a final node; the diff trace stops as soon as one AD reaches an action that cannot be matched by an action in the other AD. The set *addiff* does not include traces that have a prefix that is by itself a diff trace. In addition, in [9] we limit the results to one shortest diff trace per initial state.

Diff traces can be considered a special kind of model-based traces [8]. Each diff trace is presented to the engineer both textually and visually, by enumerating and highlighting the nodes participating in the trace on top of the concrete syntax of the input ADs themselves (see [9]).

To define a summary of *addiff*, we define $Q_l$, which partitions the set of all diff traces based on the list of actions (action names) appearing in each trace; i.e., traces that differ in terms of input or internal variable values but agree on the list of actions to be executed are considered equivalent. More formally:

**Definition 4 (action-list-equivalence partition for traces).** *The action-list-equivalent partition maps every trace $tr_1$ to the set of all traces whose list of executed actions is equal to the list of executed actions in $tr_1$: $part(tr_1) = \{tr|\ tr|_{action} = tr_1|_{action}\}$.*

For example, considering the ADs shown in Sect. 2, $addiff_{Q_l}$ of traces in $ad.v2$ that are not possible in $ad.v1$ will include (1) a trace with $tickets < 8$ where `accounts` comes immediately after `welcome msg`, and (2) one trace with $8 \leq tickets < 12$ ending with `welcome msg`. That is, the summary will include only 2 traces, each consisting of a different list of actions. However, $addiff_{Q_l}$ of traces in $ad.v3$ that are not possible in $ad.v2$ will include 6 traces, all with $tickets < 12$ and ending with the action `report`, due to the 6 possible orderings of the actions inside the fork/join (`reserve`, `accounts`, `update`).

Thus, we suggest also an alternative partition $Q_s$, where two traces are considered equivalent iff the sets of actions included in them are identical. This induces a coarser partition, as it abstracts away the order of actions in the traces. More formally:

**Definition 5 (action-set-equivalence partition for traces).** *The action-set-equivalent partition maps every trace $tr_1$ to the set of all traces whose set of executed actions is equal to the set of executed actions in $tr_1$: $part(tr_1) = \{tr|\ actions(tr) = actions(tr_1)\}$.*

Applying this coarser partition to our example, the summary for traces in $ad.v3$ that are not in $ad.v2$ includes only a single trace, where $tickets < 12$.

Finally, it is important to note that in addition to the concrete representatives, as part of the results of the computation for $addiff_{Q_l}$ and $addiff_{Q_s}$ (see below), we have symbolic representations of the initial states related to each equivalence class. These can be presented to the engineer together with the concrete traces, as part of the summary.

## 4 Computing Summaries

A naive approach to compute $diff_Q$ would first compute and enumerate all diff witnesses in $diff$ and then group them into equivalence classes according to the given partition and choose one witness from each class. This approach, however, is inefficient, as the total number of witnesses is typically an order of magnitude larger than the number of equivalence classes in the partition. Thus, a more efficient approach should be taken. We give an overview of our approach to compute $diff_Q$, for $cddiff$ and $addiff$, given the partitions suggested above.

### 4.1 Computing summaries for *cddiff*

In [11] we showed how *cddiff* can be computed (in a bounded, user-defined scope) using a translation to Alloy. Roughly, the translation takes two CDs as input and outputs an Alloy module whose instances, if any, represent object models in the semantics of one CD that are not in the semantics of the other. Computing another witness is done by asking Alloy for another instance of the module (technically, by constraining the SAT solver further to not allow the instances that were already found).

To compute $cddiff_Q$, when a diff witness is found, rather than simply asking Alloy for another witness, we generalize the instance that was found to its set of classes, and create a new predicate that specifies that it should not be the case that this set of classes consists of exactly the classes appearing in an instance of the Alloy module. We then rerun Alloy on a revised module, strengthened by the new predicate. This guarantees that a new instance, if any is found, would be a diff witness from a different equivalence class. We iterate until no more new diff witnesses are found.

The above technique is guaranteed to provide a single representative from each equivalence class without the need to enumerate all witnesses first. However, like all other analysis done with Alloy, it is bounded by a user-defined scope. Also, its performance may not scale well for large CDs. Addressing these limitations may require the use of a completely different solution, i.e., not using Alloy, and is left for future work.

### 4.2 Computing summaries for *addiff*

In [9] we showed how *addiff* can be efficiently computed using a symbolic fixpoint algorithm, based on BDDs and the technologies of symbolic model-checking [2]. The algorithm starts with a representation of all non-corresponding states. It then moves 'backward', and adds to the current set of states, states from which there exists a successor in one AD such that for all successors in the other AD, the resulting successor pair is in the current set of states. The steps 'backward' continue until reaching a least fixpoint, i.e., until no more states are added. When the fixpoint is reached, the algorithm checks whether the fixpoint set includes initial states. For each such initial state, if any, the algorithm uses the sets of states computed during the backward steps to move forward (from the minimal position it can start from) and construct shortest diff traces.

To compute $addiff_{Q_l}$ we start with the first phase of the original algorithm and symbolically compute the set of all initial states from which a diff trace may start and all sets of states included in all diff traces. Then, rather then enumerating all concrete diff traces by computing a concrete diff trace starting in each initial state, we start with the set of all initial states and symbolically move forward to the set of all next states. If two or more actions are possible in the next step, we split the set of next states according to their action and continue, symbolically, for each of the sets in the split. We iterate this until reaching the differentiating actions, i.e., until no corresponding next state exists. Finally, for each symbolic trace we now have, we choose one initial state and compute a concrete trace that starts from it. We symbolically represent the set of initial states that share the list of actions in the trace (e.g., with ranges of input variables).

Computing a summary with our coarser partition, $addiff_{Q_s}$, is similar. When we are done with computing the symbolic traces of the action-list partition, before choosing concrete representatives, we iterate over the set of symbolic traces and eliminate any symbolic trace whose set of actions already appeared (in another order) in a previous trace. For each of the remaining symbolic traces, we choose one initial state and compute a concrete diff trace that starts from it.

## 5 Initial Evaluation and Discussion

We have applied the above summarization strategies for *cddiff* and *addiff* to the examples of Sect. 2. Table 1 lists the results in terms of the number of diff witnesses (object models, traces) found, with and without summarization.

For *cddiff*, all our examples have 20 or more diff witnesses without summarization (we computed *cddiff* with a scope of 10 and stopped after finding 20 witnesses). The number of witnesses found with summarization was only 3 or 4. The results show the effectiveness of the summarization approach in significantly reducing the number of diff witnesses presented to the engineer while keeping the set as diverse as possible. Also, note that finding only 3 witnesses means that the SAT solver was executed only 4 times (in the last execution, no diff witness was found). This shows the efficiency of our approach.

For *addiff*, the number of diff traces found without summarization varied: for some examples there are only few diff traces, while for others the number of diff traces found was much higher, up to 72. Applying summarization to the examples with a small number of witnesses does not make much difference. However, applying summarization to the examples with the many witnesses results in significantly smaller sets of witnesses, up to at most 6 representative traces for each example. For the action-list partition, significant reduction is observed when the ADs state space is large due to many possible inputs (many variables or variables with large domains like our *tickets* variable). For the action-set partition, further reduction is observed when the ADs' state space is large and where differences occur after some fork/join blocks with much partial order.

In the general case, summarization may entail information loss: one cannot always use the summary to enumerate all witnesses. Yet, in some cases, it is possible to keep an efficient symbolic representation of each equivalence class within the summary, so that *diff* can be easily computed from $diff_Q$. For example, the computation of $addiff_{Q_l}$, based on the action-list partition, includes a symbolic representation of all input states where diff traces may start. Given initial states and the list of actions that characterize each of the partitions, all diff traces can be reconstructed. For $addiff_{Q_s}$, based on the action-set partition, however, this is not the case; once the order of actions is abstracted away, one cannot use an initial state to generate a trace that is guaranteed to be a diff trace.

Finally, we consider the following alternatives for semantic differencing summarization. First, one may suggest a partition based on syntactic differences, i.e., such that witnesses are classified according to the syntactic differences they 'cover'. Second, in addition to partitioning, one may be interested in defining a (partial) order, such that the summarization method chooses a representative that is also minimal within its equivalence class. For example, in the case of *cddiff*, a partial order may be defined based on diff witness size, i.e., the number of objects in the object model. In the case of *addiff*, a partial order may be defined based on diff traces length. It seems that smaller diff witnesses would be easier to present and understand. More generally, rather than ordering witnesses locally, within each equivalence class, one may suggest a pre-order on all diff witnesses and look for global minimal ones.

| Name | # Wit. found | # Wit. found with summarization |
|---|---:|---:|
| $cd.v1$ vs. $cd.v2$ | 20 | 3 |
| $cd.v2$ vs. $cd.v1$ | 20 | 4 |
| $ad.v1$ vs. $ad.v2$ | 4 | 1/1 |
| $ad.v2$ vs. $ad.v1$ | 20 | 3/3 |
| $ad.v2$ vs. $ad.v3$ | 72 | 6/1 |
| $ad.v3$ vs. $ad.v2$ | 72 | 6/1 |
| $ad.v1$ vs. $ad.v3$ | 28 | 4/2 |
| $ad.v3$ vs. $ad.v1$ | 36 | 5/4 |

**Table 1.** Results of applying the summarization strategies to the examples from Sect. 2. We computed CDDiff with scope 10 and stopped after 20 witnesses were found. For ADDiff summarization we show the number of witnesses according to the action-list partition / action-set partition.

Formalizing and evaluating these alternatives is left for future work.

## 6  Related Work

The problem of summarizing semantic differences is close to the problem of effective design space exploration [7], where the goal is to quickly visit a diverse set of solutions across a design space. The approach in [7] takes a user-defined notion of equivalence as input, and generates symmetry breaking predicates, which ensure that the underlying exploration engine does not sample multiple equivalent design candidates. In addition, the work employs randomization to incrementally construct a diverse set of non-isomorphic solutions, ideally making the solver 'jump around' various parts of the design space, sampling a wide variety of solutions. The work is integrated in a tool called FORMULA, which uses an SMT solver.

In [6], the authors present a technique to summarize all counterexamples of an LTL model-checking problem. They generalize concrete examples found by the SMV model-checker into equivalence classes, describe these with LTL formulas, and re-run the model-checker on a revised formula where examples that are equivalent to previously found ones are not considered. The work suggests four specific kinds of equivalence, at different levels of abstraction.

Our work is similar, in that we use class equivalence as a criteria for results selection and presentation. It is also very different, as it is specifically applied to the problem of semantic differencing for several modeling languages, and thus the criteria for 'symmetry breaking' and the technologies used (Alloy/SAT, BDD-based algorithms) are specific and very different than the ones in [6, 7].

Many works present syntactic approaches to differencing (e.g., [1, 13, ?]). Some related tools support hierarchical presentation of differences where the hierarchy is defined by the abstract syntax tree (AST) [3]. All differences are computed but the presentation in the AST can encapsulate them under collapsed sub-trees. This may be viewed as a form of presentation summarization. Note that our summarization technique is not limited to the presentation but is applied already as part of the computation: we show how to compute the summary set without the enumeration of all witnesses during the computation.

We are not aware of other work in the domain of software evolution that is directly related to differences summarization.

## 7 Conclusion

We have presented summarization techniques for semantic model differencing. We motivated the challenge of summarization and suggested ways to address it in the context of CD and AD semantic differencing. We demonstrated the utility of our summarization approach in providing a small yet informative set of diff witnesses and discussed alternatives and future challenges.

Future work includes the integration of the techniques presented here into the prototype implementations presented in [9] and [11]. Moreover, as we extend semantic differencing to additional languages, e.g., feature models and statecharts, we will be looking for summarization techniques for these languages too.